\documentclass[aps,prd,onecolumn,groupedaddress,showpacs,nofootinbib,amssymb]{revtex4}
\usepackage{amssymb}

\usepackage{graphicx,bm}
\usepackage{hyperref}

\usepackage[english]{babel}

\usepackage{amsmath}

\usepackage{amsfonts}
\usepackage{subfig}

\usepackage{epsfig}

\def\nn{\nonumber \\}

\newcommand{\be}{\begin{equation}}

\newcommand{\ee}{\end{equation}}

\newcommand{\bea}{\begin{eqnarray}}

\newcommand{\eea}{\end{eqnarray}}

\newcommand{\beq}{\begin{eqnarray}}

\newcommand{\eeq}{\end{eqnarray}}

\begin{document}
\title{$f(R,T,R_{\mu\nu}T^{\mu\nu})$  gravity phenomenology and $\Lambda$CDM universe}
\author{Sergei D. Odintsov$^{(a,b)}$\footnote{odintsov@ieec.uab.es, also at Tomsk State Pedagogical University, Tomsk, Russia and ENU, Astana} and Diego S\'{a}ez-G\'{o}mez$^{(c,d)}$\footnote{diego.saezgomez@uct.ac.za} }
\affiliation{$^a$Instituci\`{o} Catalana de Recerca i Estudis Avan\c{c}ats
(ICREA), Barcelona, Spain \\
$^b$Institut de Ci\`encies de l'Espai
ICE (CSIC-IEEC), Campus UAB Facultat de Ci\`encies, Torre C5-Parell-2a
pl, E-08193 Bellaterra (Barcelona) Spain \\
$^c$\ Astrophysics, Cosmology and Gravity Centre (ACGC) and \\ 
Department of Mathematics and Applied Mathematics, University of Cape Town, Rondebosch 7701, Cape Town, South Africa \\
$^d$\ Fisika Teorikoaren eta Zientziaren Historia Saila, Zientzia eta Teknologia Fakultatea,\\
Euskal Herriko Unibertsitatea, 644 Posta Kutxatila, 48080 Bilbao, Spain}

\begin{abstract}

We propose general $f(R,T,R_{\mu\nu}T^{\mu\nu})$ theory as generalization of covariant Ho\v{r}ava-like gravity with dynamical Lorentz symmetry breaking. FRLW cosmological dynamics for several versions of such theory is considered. The reconstruction of the above action is explicitly done, including the numerical reconstruction for the occurrence of $\Lambda$CDM universe.
De Sitter universe solutions in the presence of non-constant fluid are also presented. The problem of matter instability in $f(R,T,R_{\mu\nu}T^{\mu\nu})$ gravity is discussed.
\end{abstract}


\maketitle

%%%%%%%%%%%%%%%%%%%%%%%%%%%
\section{Introduction}
%%%%%%%%%%%%%%%%%%%%%%%%%%%

Recently, much attention has been paid on modified gravities, and specifically on their possible role in the accelerated expansion of the universe, a fact widely accepted by a major part of the scientific community, and strongly supported by the observations.  In this sense, the so-called $f(R)$ gravity is the easiest and most popular classical extension of General Relativity (GR) because of its simplicity and absence of ghosts (for a recent review on $f(R)$ gravity, see Ref.~\cite{0601213}). In addition, the reconstruction of $f(R)$ theories that are capable of reproducing the dark energy epoch, and even the inflationary phase, is generally straightforward in comparison with other modifications of gravity (see Refs.~\cite{Ref5,gr-qc/0607118,Nojiri:2003ft}). $f(R)$ gravity is able of reproducing $\Lambda$CDM epoch (see Ref.~\cite{gr-qc/0607118}), or mimicking  a cosmological constant at the current era, and even unifying the entire cosmological history \cite{Nojiri:2003ft}. Moreover, some $f(R)$ gravities, so-called {\it viable} models, can pass the local gravitational tests and reproduce a realistic cosmological evolution (see Ref.~\cite{f(R)viable}). As well viable $f(R)$ gravity may pass the matter instability  \cite{Dolgov:2003px}, a large perturbation that may occur in the study of celestial body solution. Nevertheless, viable $f(R)$ models are known to reproduce a kind of future singularity (see Ref.~\cite{Appleby:2009uf}), a problem that can be circumvented in the scalar-tensor representation of $f(R)$ gravity, \cite{Saez-Gomez3}. Other  modified gravities (like Gauss-Bonnet ones)
 which are also capable of reproducing the dark energy epoch have been also studied \cite{Cog1}. \\

However, much less attention has been paid to more complex theories, specially to those theories that contain a non-standard coupling of curvature with energy-momentum tensor. Recently,  class of modified gravity theories in which the gravitational action contains a general function $f(R,T)$, where $R$ and $T$ denote the curvature and the trace of the energy-momentum tensor respectively, has been proposed (see Ref.~\cite{Harko:2011kv}). Cosmological evolution of such theories has been studied, including  the reconstruction of cosmological solutions and the presence of future singularities (see Ref.~\cite{stephaneseul3}). Nevertheless, such strong coupling of the curvature and the trace $T$ implies the violation of the usual continuity equation, an issue that can be solved by an appropriate function $f(R,T)$, as shown for the first time in Ref.~\cite{perturbationsfrt}. Nonetheless, this function gives rise to an evolution of cosmological perturbation for the sub-Hubble modes which deviates significantly from the standard GR results, eventually leading to singularities of matter perturbations (see Ref.~\cite{perturbationsfrt}) and to the appearance of an extra force in geodesic equation if conservation law is violated. \\

In the present work, an extension of $f(R,T)$ gravity is proposed. The action described by a generic function $f(R,T,R_{\mu\nu}T^{\mu\nu})$ is considered, where $R$ is the scalar curvature, $T$ is the trace of the energy-momentum tensor, and $R_{\mu\nu}$ is the Ricci tensor. Then, the general FRLW field equations are derived in the presence of $R_{\mu\nu}T^{\mu\nu}$ coupling terms. Several cosmological solutions are studied, where it is found that generally the matter sector does not behave as in GR, since the divergence of the field equations does not lead to the usual continuity equation. Nevertheless, this issue can be solved by assuming a particular $f(R,T,R_{\mu\nu}T^{\mu\nu})$, where due to the extra degree of freedom, the usual cosmological evolution for a perfect fluid can be imposed. In addition, a pure de Sitter solution is discussed, which can be realized even in the presence of a non-constant fluid. The occurrence of $\Lambda$CDM universe is also analyzed: the corresponding gravitational action $f(R,T,R_{\mu\nu}T^{\mu\nu})$ is reconstructed, where GR with a cosmological constant is just a particular solution. Finally, the matter instability, a perturbation that may be induced in the interior of celestial body solution, is discussed for such class of models.\\

Note that physical motivation for the theory under consideration comes from covariant Ho\v{r}ava-like gravity with dynamical breaking of Lorentz invariance. Actually, such power-counting renormalizable covariant gravity \cite{cov} represents simplest, power-law version of $f(R,T,R_{\mu\nu}T^{\mu\nu})$ theory. One can expect that such theories may provide deeper connection between modified gravity and Ho\v{r}ava-Lifshitz theory.

%%%%%%%%%%%%%%%%%%%%%%%%%
\section{$f(R,T,R_{\mu\nu}T^{\mu\nu})$ gravity}\label{General}
%%%%%%%%%%%%%%%%%%%%%%%%%

Let us start from the general action for $f(R,T,R_{\mu\nu}T^{\mu\nu})$ gravity,
\be
S=\frac{1}{2\kappa^2}\int d^4x\sqrt{-g}\ f(R,T,R_{\mu\nu}T^{\mu\nu})+\int d^4x\sqrt{-g}\ \mathcal{L}_m=S_G+S_m\ .
\label{I1}
\ee
Here, $\kappa^2$ is gravitational coupling constant,  $R$ is the Ricci scalar and $T$ represents the trace of the energy-momentum tensor, $T=T^{\mu}_{\mu}$, while $\mathcal{L}_m$ is the matter Lagrangian. As usually the energy-momentum tensor is defined as,
\be
T^{\mu\nu}=\frac{2}{\sqrt{-g}}\frac{\delta S_m}{\delta g_{\mu\nu}} .
\label{I2}
\ee
Then, by assuming a matter action that only depends on the metric but does not on its first derivatives, the energy-momentum tensor yields,
\be
T^{\mu\nu}=g^{\mu\nu}\mathcal{L}_m+2\frac{\delta\mathcal{L}_m}{\delta g_{\mu\nu}}\ .
\label{I2a}
\ee
Hence, field equations are obtained by varying the action (\ref{I1}) with respect to the metric $g^{\mu\nu}$,
\[
\delta S=\frac{1}{2\kappa^2}\int d^4x \left[\delta(\sqrt{-g})f(R,T,R_{\mu\nu}T^{\mu\nu})+\sqrt{-g}\left(f_R \delta R+f_T \delta T+f_P \delta R_{\mu\nu}T^{\mu\nu}\right)\right]
\]
\be
+\int d^4x \delta \left(\sqrt{-g}\mathcal{L}_m\right)\ .
\label{I3}
\ee
Here $P=R_{\mu\nu}T^{\mu\nu}$ has been defined for convenience and in order to simplify the expressions that are coming below, whereas the subscript means variation with respect to $P$, $R$ and $T$. Those terms corresponding to the variations of the metric determinant, the Ricci scalar and  the trace of the energy-momentum tensor have been already found in the literature (see Ref.~\cite{Harko:2011kv})
\bea
\delta\sqrt{-g}=-\frac{1}{2}\sqrt{-g}\ g_{\mu\nu}\delta g^{\mu\nu}\ , \label{I3b}
 \\
f_R \delta R+f_T \delta T=\left[R_{\mu\nu} f_R+(g_{\mu\nu}\Box-\nabla_{\mu}\nabla_{\nu})f_R+\left(T_{\mu\nu}+\Theta_{\mu\nu}\right)f_T\right] \delta g^{\mu\nu}\ .
\label{I3b}
\eea
where the second term in the rhs of (\ref{I3b}) has been integrated by parts, whereas the tensor  $\Theta_{\mu\nu}$ is defined as
\be
\Theta_{\mu\nu}=g^{\alpha\beta}\frac{\delta T_{\alpha\beta}}{\delta g^{\mu\nu}}=g_{\mu\nu}\mathcal{L}_m-2T_{\mu\nu}-2g^{\alpha\beta}\frac{\delta^2 \mathcal{L}_m}{\delta g^{\mu\nu}\delta g^{\alpha\beta}}\ .
\label{I4}
\ee
The terms corresponding to the variation of $R_{\mu\nu}T^{\mu\nu}$ yield
\be
f_P \delta (R_{\mu\nu}T^{\mu\nu})=f_P\left(\delta R_{\mu\nu} T^{\mu\nu}+R_{\mu\nu} \delta T^{\mu\nu}\right)\ .
\label{I4a}
\ee
Since the variation of the Ricci tensor  is given by
\be
\delta R_{\mu\nu}=\nabla_{\sigma}\delta\Gamma^{\sigma}_{\mu\nu}-\nabla_{\nu}\delta\Gamma^{\sigma}_{\mu\sigma}\ ,
\label{I4b}
\ee
where
\be
\delta\Gamma^{\sigma}_{\mu\nu}=\frac{1}{2}g^{\sigma\lambda}\left(\delta g_{\mu\lambda;\nu}+\delta g_{\nu\lambda;\mu}-\delta g_{\mu\nu;\lambda}\right)\ .
\label{I4c}
\ee
The variation of the first term in (\ref{I4a}) with respect to the metric yields
\be
f_PT^{\mu\nu}\delta R_{\mu\nu}=\left[\frac{1}{2}\left(\Box f_PT_{\mu\nu}+g_{\mu\nu}\nabla_{\alpha}\nabla_{\beta}f_PT^{\alpha\beta}\right)-\nabla_{\alpha}\nabla_{\nu}f_PT^{\alpha}_{\mu}\right]\delta g^{\mu\nu}\ ,
\label{I4d}
\ee
whereas the variation over the second term in (\ref{I4a}) is given by,
\be
f_PR_{\mu\nu}\delta T^{\mu\nu}=f_P\left[-G_{\mu\nu}\mathcal{L}_m-\frac{1}{2}RT_{\mu\nu}+2R^{\alpha}_{\mu}T_{\alpha\nu}-2R^{\alpha\beta}\frac{\delta^2 \mathcal{L}_m}{\delta g^{\mu\nu}\delta g^{\alpha\beta}}\right]\delta g^{\mu\nu}
\label{I4e}
\ee
Note that for a regular $f(R,T,R_{\mu\nu}T^{\mu\nu})$ function, in the absence of any kind of matter, the  $f(R)$ gravity equations are recovered. The corresponding properties and solutions well studied in the literature on $f(R)$ gravity are also satisfied by $f(R,T,R_{\mu\nu}T^{\mu\nu})$ theories in classical vacuum (for a recent review on $f(R)$ theories, see \cite{0601213}). Moreover, here we are interesting to study the behavior of this kind of theories for flat FLRW metrics, which  are expressed in comoving  coordinates by the line element
\be
ds^2=-dt^2+a(t)^2\left(dr^2+r^2d\theta^2+r^2\sin^2\theta\ d\varphi^2\right)\ ,
\label{I5}
\ee
where $a(t)$ is the scale factor. Then, the main issue arises on the content of the Universe through the energy-momentum tensor,  and consequently on the matter Lagrangian $\mathcal{L}_m$ through the terms $R_{\mu\nu}T^{\mu\nu}$ and $T=T^{\mu}_{\mu}$. Since the flat FLRW cosmology (\ref{I5}) is assumed, the content of the universe (pressureless matter, radiation,... ) can be well described by a perfect fluid, whose energy-momentum tensor can be defined in several ways, but which is assumed here to have the form,
\be
T_{\mu\nu}=pg_{\mu\nu}+(\rho+p)u_{\mu}u_{\nu}\ .
\label{I6}
\ee
Here $\rho$ and $p$ are the energy and pressure densities respectively, and $u^{\mu}$ is the four-velocity of the fluid, which satisfies $u_{\mu}u^{\mu}=-1$, and is given by $u^{\mu}=(1\ 0\ 0\ 0)$ in comoving coordinates. The definition of the matter Lagrangian for a perfect fluid is not unique, following the definition suggested in Ref.~\cite{Harko:2011kv}, and in order to be consistent with the variation of the energy-momentum-tensor (\ref{I6}) with respect to the metric, $\mathcal{L}_m=p$ is assumed, and consequently the second variation of the matter Lagrangian in (\ref{I4}) and (\ref{I4e}) becomes null. However, note that other choices of the matter Lagrangian would give rise to the same results. Hence, the tensor $\Theta_{\mu\nu}$, defined in (\ref{I4}), yields,
 \be
\Theta_{\mu\nu}= -2 T_{\mu\nu}+p\ g_{\mu\nu}\ ,
\label{I7}
\ee
whereas the tensor $\Xi_{\mu\nu}$ defined in (\ref{I4e}) can be expressed as,
\be
\Xi_{\mu\nu}=-G_{\mu\nu}p-\frac{1}{2}RT_{\mu\nu}+2R^{\alpha}_{\;\;(\mu}T_{\nu)\alpha}\ ,
\label{I7b}
\ee
Hence, the complete set of the field equations is given by,
\[
R_{\mu\nu}f_R-\frac{1}{2}g_{\mu\nu}f+\left(g_{\mu\nu}\Box-\nabla_{\mu}\nabla_{\nu}\right)f_R+\left(T_{\mu\nu}+\Theta_{\mu\nu}\right)f_T
\]
\be
+\frac{1}{2}\left(\Box T_{\mu\nu}f_P+g_{\mu\nu}\nabla_{\alpha}\nabla_{\beta}T^{\alpha\beta}f_P\right)-\nabla_{\alpha}\nabla_{(\mu}T^{\alpha}_{\nu)}f_P+\Xi_{\mu\nu}f_P=\kappa^2T_{\mu\nu}\ ,
\label{I7c}
\ee
where recall that $P=R_{\mu\nu}T^{\mu\nu}$. Then, for a particular equation of state (EoS) $p=w\rho$, the corresponding FLRW equations can be obtained, and its cosmology studied. Nevertheless, note that the usual continuity equation is not satisfied in general, since the divergence of the equation (\ref{I7c}) is not null. In the case of $f(R,T)$ gravity, a particular form of the action that recovers the continuity equation was found in Ref.~\cite{perturbationsfrt}. Nevertheless, in the case of $f(R,T,R_{\mu\nu}T^{\mu\nu})$ gravity, where terms proportional to $R_{\mu\nu}T^{\mu\nu}$ appear in the action, an explicit form of the gravitational Lagrangian that satisfies the continuity equation can not be obtained in general, since the equation involves  very complex expressions that does not allow to get an explicit form of the action $f(R,T,R_{\mu\nu}T^{\mu\nu})$, as pointed out in Ref.~\cite{T.Harko}. In addition, note that the presence of different species in the matter Lagrangian may lead to different constraints in the action, for instance the electromagnetic field, where $T=0$. Nevertheless, by exploring some particular cosmological solutions, the dynamics of the matter sector may lead to similar behaviors as in GR.

%%%%%%%%%%%%%%%%%%%%%%%%%%%%%%%%%
\section{Cosmological solutions} \label{Evolution}
%%%%%%%%%%%%%%%%%%%%%%%%%%%%%%%%%

In this section, some particular cosmological solutions are studied, and the corresponding action is reconstructed. A pressureless fluid, where $w_m=0$, is assumed along the section. From one side such choice simplifies the calculations, and from the other side, a presureless fluid represents a suitable description of the baryonic and cold dark matter content of the universe, whereas the dark energy sector is comprised by the modifications of GR. Then, the energy-momentum tensor reads $T^{\mu}_{\nu}=\text{Diag}\left(-\rho, 0, 0, 0\right)$, whereas the trace $T$ and the scalar $P$ give rise to
\be
T=T^\mu_\mu=-\rho\ , \quad P=R_{\mu\nu}T^{\mu\nu}=R_{00}T^{00}=-3 \rho (H^2+\dot{H})\ .
\label{2.1}
\ee

%%%%%%%%%%%%%%%%%%%%%%%%%%%%%%%%%%%%%%%%%%%%%
\subsection{Model $f(R,R_{\mu\nu}T^{\mu\nu})=\alpha R+f(R_{\mu\nu}T^{\mu\nu})$}
%%%%%%%%%%%%%%%%%%%%%%%%%%%%%%%%%%%%%%%%%%%%%

Let us start by considering the action $f(R,R_{\mu\nu}T^{\mu\nu})=\alpha R+f(R_{\mu\nu}T^{\mu\nu})$, where the dependence on the trace $T$ is omitted. By assuming the flat FLRW metric (\ref{I5}), the FLRW equations are \bea
3\alpha H^2+\frac{1}{2}\left[f-3H\partial_t(\rho f_P)-3\left(3H^2-\dot{H}\right)\rho f_P\right]-\kappa^2 \rho=0\ , \nn
-\alpha\left(3H^2+2\dot{H}\right)-\frac{1}{2}\left[f-\partial_{tt}(\rho f_P)-4H\partial_t(\rho f_P)-\left(3H^2+\dot{H} \right)\rho f_P\right]=0\ .
\label{2.2}
\eea
At this point, a particular cosmological solution, described by a given Hubble parameter $H(t)$ can be considered. The corresponding action $f(R_{\mu\nu}T^{\mu\nu})$ could be reconstructed. In addition, one can note that the usual continuity equation is not longer valid. Hence, the time dependence of the energy-density $\rho(t)$ has to be also obtained from the above equations (\ref{2.2}), which implies more complicated treatment for reconstructing the gravitational theory than within the framework of $f(R)$ gravity. \\

First of all, let us consider cosmological solutions of the type,
\be
a(t)=a_0 t^{m}\ \quad \rightarrow \quad H=\frac{m}{t}\ ,
\label{2.3}
\ee
that shall be refereed to as power-law behavior. Within GR, this type of solutions accomplishes the scale factor evolution for perfect fluids with a constant EoS, such as dust ($m=2/3$) or radiation ($m = 1/2$) dominated
Universe. Then, since the FLRW equations (\ref{2.2}) shall be composed by powers of the time variable, a natural consideration would be to assume $\rho(t)$ and $f(P(t))$ to be powers of the cosmic time as well,
\be
\rho(t)=\rho_0 t^{\sigma}\ , \quad  f=\beta P^n=\beta\left(R_{\mu\nu}T^{\mu\nu}\right)^n= \beta \left[-3\rho_0m(m-1)\right]^n t^{n(\sigma-2)}\ ,
\label{2.4}
\ee
Then, the FLRW equations (\ref{2.2}) yield
\[
3\alpha\frac{m^2}{t^2}-\kappa^2 \rho_0 t^{\sigma}+\beta \frac{\left[n^2(\sigma-2)+3n(m+1)+m-1\right]\left[-3m(m-1)\rho_0\right]^{n}}{2(m-1)}t^{n(\sigma-2)}=0\ ,
\]
\[
\frac{6\alpha (m-1)(3m-2)m^2}{6 (m-1)m t^2}+\beta\left[3(m-1)m+n(2+m)(1+3m)\right.
\]
\[
\left.+n^2(\sigma-2)(4m-3)+n^3(\sigma-2)^2\right]/6 (m-1)m
\]
\be
\times\left[-3\rho_0m(m-1)\right]^n t^{n(\sigma-2)}=0
\label{2.5}
\ee
In general, the matter density does not satisfy the continuity equation, but according to (\ref{2.3}) and (\ref{2.4}), if $\sigma=-3m$ the usual evolution for a pressureless fluid is recovered. Let us consider several cases depending on the choice of the coupling constants.

\begin{itemize}
\item In the most general case, where all the coupling constants are non-zero, the only possible solution of the equations yields,
\be
\sigma=-2\ , \quad n=1/2\ .
\label{2.6}
\ee
Whereas, $\{\alpha,\beta\}$ are expressed in terms of $m$,
\[
\alpha=\frac{9m-7}{6m(3m^2-3m+1)}\kappa^2\rho_0\ , 
\]
\be
 \beta=\frac{2(3m-2)(m-1)}{(3m^2-3m+1)\sqrt{-3m(m-1)}}\kappa^2 \sqrt{\rho_0}\ .
\label{2.7}
\ee
In order to ensure a real and physical gravitational action, $m<1$ according to the expression of $\beta$ in (\ref{2.7}), which allows the radiation or matter dominated evolution of GR-type, but forbids any accelerating solution ($m>1$). Moreover, for  $m=2/3$, the usual continuity equation is recovered, which leads to a dust matter-like dominated evolution. Hence, the above gravitational theory is capable of reproducing the matter dominated epoch with the presence of the extra terms in the action.
 
\item By considering $\alpha=0$, the gravitational action turns out to be $f=\beta P^n=\beta (R_{\mu\nu}T^{\mu\nu})^n$, and from (\ref{2.5}),
\be
n=\frac{\sigma}{\sigma-2}\ .
\label{2.8}
\ee
Then, the equations (\ref{2.5}) give rise to a system of two transcendental equations that can be solved numerically, with three variables $\{m,\sigma,\beta\}$ that lead to an infinite number of solutions. By setting $\sigma=-3m$ in order to satisfy the continuity equation, the number of variables is reduced and the system can be solved numerically. 
\item Finally, by assuming that the coupling among the gravitational and matter sectors is described solely through $R_{\mu\nu}T^{\mu\nu}$, basically by setting $\kappa^2=0$, the exponent of the action (\ref{2.4}) is constrained to,
\be
n=-\frac{2}{\sigma-2}\ .
\label{2.9}
\ee
Then, as above, the equations (\ref{2.5}) become a system of algebraic equations for the constants $\{m,\sigma\}$ in terms of $\{\alpha,\beta\}$, which describes a set of infinite solutions, but confirms the suitability of the action with $f(P)=P^n=(R_{\mu\nu}T^{\mu\nu})^n$ for reproducing power-law expansion. As above the number of variables of the equations can be reduced by imposing $\sigma=-3m$.
\end{itemize}´
Nevertheless, the reconstruction of the exact expression for the gravitational action is not possible in general, and numerical calculations are required, as shown below. In addition, note that the evolution of the energy-density behaves differently to the one given in GR. This may not be important for the illustration of the reconstruction procedure, as for instance, the solutions considered above, but it may seriously affect  the observational constraints when a realistic evolution is studied. In this sense, the FLRW equations (\ref{2.2})  are two differential equations for the functions $\rho(z)$ and $f(R_{\mu\nu}T^{\mu\nu})=f(z)$, expressed in terms of the redshift. They do not ensure that the solution for the energy-density behaves in the proper way as observations suggest, namely $\rho_m\propto(1+z)^3$ for dust matter, apart from the trivial solution $f(R_{\mu\nu}T^{\mu\nu})=-2\Lambda$ that recovers GR with a cosmological constant, and the usual continuity equation is satisfied again. Hence, more general actions should be considered in order to provide a realistic evolution.

%%%%%%%%%%%%%%%%%%%%%%%%%%%%%%%%%%%%%%%%%%%%
\subsection{Model $f(R,T,R_{\mu\nu}T^{\mu\nu})= R+f(R_{\mu\nu}T^{\mu\nu})+g(T)$}
%%%%%%%%%%%%%%%%%%%%%%%%%%%%%%%%%%%%%%%%%%%%

By assuming an action described by $f(R,T,R_{\mu\nu}T^{\mu\nu})=R+f(R_{\mu\nu}T^{\mu\nu})+g(T)$, the FLRW equations are slightly modified in comparison with the above case,
\bea
3 H^2+\frac{1}{2}\left[f+g-3H\partial_t(\rho f_P)-3\left(3H^2-\dot{H}\right)\rho f_P\right]- \rho (\kappa^2-g_T)=0\ , \nn
-3H^2-2\dot{H}-\frac{1}{2}\left[f+g-\partial_{tt}(\rho f_P)-4H\partial_t(\rho f_P)-\left(3H^2+\dot{H} \right)\rho f_P\right]=0\ .
\label{2.10}
\eea
Within this section, the above equations are expressed in terms of the redshift $1+z=\frac{1}{a(t)}$, where $\frac{d}{dt}=-(1+z)H\frac{d}{dz}$, whereas $P=P(z)$ and $T=T(z)$. In addition, a new variable is defined $\chi(z)=\rho f_P$ which simplifies the equations (\ref{2.10}) by extending the system to a set of three differential equations
\[
3H^2+\frac{1}{2}\left\{f(z)+g(z)-3\left[3H^2+(1+z)H H'\right] \chi(z)+3H^2(1+z)\chi'(z)\right\}
 \]
 \[
 -\left(\kappa^2+\frac{g'(z)}{\rho'(z)}\right)\rho(z)=0\ ,
 \]
\[
-3H^2+(1+z)HH'-\frac{1}{2}\left\{f(z)+g(z)-\left[3H^2-(1+z)HH'\right]\chi(z)+ \right.
\]
\[
\left.\left[3(1+z)H^2-(1+z)^2HH'\right]\chi'(z) +(1+z)^2 \chi''(z)\right\}=0\ , 
\]
\be
\chi(z)-\frac{\rho(z)f'(z)}{P'(z)}=0\ .
\label{2.11}
\ee

Hence, the system is composed by three equations with four unknown functions $\{f(z),g(z),\chi(z),\rho(z)\}$. In comparison with the above section, the presence of an additional function $g(T)$  introduces a new degree of freedom that allows to impose the usual evolution of the energy-density
\be
\rho=\rho_0 a^{-3(1+w_m)}=\rho_0 (1+z)^{3(1+w_m)}\ ,
\label{2.12}
\ee
Recall that the reconstruction procedure is restricted to pressureless fluids in this section, so (\ref{2.12}) reduces to $\rho\propto (1+z)^3$ when dust matter $w_m=0$ is assumed. Hence, the system of equations (\ref{2.11}) is completely solved by the set of functions $\{f(z),g(z),\chi(z)\}$. In addition, note that the equations (\ref{2.11}) can be reduced to a single fourth order equation in $f(z)$ by replacing the second and third equations into the first one, whereas by assuming (\ref{2.12}), the expressions (\ref{2.1}) yield
\[
T=T^\mu_\mu=-\frac{3H_0^2}{\kappa^2}\Omega_m^0 (1+z)^{3}\ ,  
\]
\be
P=R_{\mu\nu}T^{\mu\nu}=R_{00}T^{00}=-3 \frac{3H_0^2}{\kappa^2}\Omega_m^0 (1+z)^{3} (H^2-(1+z)H H'))\ .
\label{2.12a}
\ee
where $\Omega_m^0=\frac{\rho_0}{\frac{3}{\kappa^2}H_0^2}$, and the subscript $_0$ refers to the value at $z=z_0$. Moreover, in order to keep the correct units  in the action, the functions $f(P)$ and $g(T)$ can be defined more conveniently as,
\be
f(P)=H_0^2\ F\left(\frac{P}{P_0}\right)\ , \quad g(T)=H_0^2\ G\left(\frac{T}{T_0}\right)\ ,
\label{2.16}
\ee
where
\be
P_0=-\frac{9H_0^4\Omega_m^0}{\kappa^2}\ , \quad T_0=-\frac{3H_0^2}{\kappa^2}\Omega_m^0 \ .
\label{2.17}
\ee
As an illustrative examoles, let us consider de Sitter (dS) solutions
\be
H(z)=H_0\ .
\label{dSa}
\ee
The system of equations (\ref{2.11}) can be easily reduced to a single equation in terms of $F(z)$ that yields,
\be
2(1+z)^3F^{(4)}(z)+(1+z)^2F^{(3}(z)-10(1+z)F''(z)+36F'(z)-162\Omega_m(1+z)^2=0\ ,
\label{dSb}
\ee
which is straightforward to solve. The solution is obtained in terms of the redshift,
\[
F(z)=C_1(1+z)^{\alpha}+(1+z)^{\beta}\left\lbrace C_2\cos\left[\omega \ln(1+z)\right]+C_3\sin\left[\omega \ln(1+z)\right]\right\rbrace
\]
\be
+C_4+3\Omega_m^0(1+z)^3\ ,
\label{dSc}
\ee
where
\be
\alpha=-1.327\ , \quad \beta=3.414\ , \quad \omega=1.38\ ,
\label{dSd}
\ee
whereas $C_i$ are integration constants. Moreover, for the dS solution (\ref{dSa}) and the energy-density (\ref{2.12}), the terms $P/P_0=T/T_0=(1+z)^3$, and the function $F(R_{\mu\nu}T^{\mu\nu})$ yields,
\[
F(P)=C_1\left(\frac{P}{P_0}\right)^{\alpha}+\left(\frac{P}{P_0}\right)^{\beta/3}\left\lbrace C_2\cos\left(\frac{\omega}{3} \ln\frac{P}{P_0}\right)+C_3\sin\left(\frac{\omega}{3} \ln\frac{P}{P_0}\right)\right\rbrace
\]
\be
+C_4+3\Omega_m^0\frac{P}{P_0}\ .
\label{dSe}
\ee
Whereas $G(T)$ acquires a similar form
\[
G(T)= \tilde{C}_1\left(\frac{T}{T_0}\right)^{\alpha/3}+ \left(\frac{T}{T_0}\right)^{\beta/3}\left\lbrace \tilde{C}_2\cos\left(\frac{\omega}{3} \ln\frac{T}{T_0}\right)+\tilde{C}_3\sin\left(\frac{\omega}{3} \ln\frac{T}{T_0}\right)\right\rbrace
\]
\be
+\tilde{C}_4-3\Omega_m^0\frac{T}{T_0}\ ,
\label{dSf}
\ee
where $\tilde{C}_i$ are constants that depend on the integration constants $C_i$ and the parameters (\ref{dSd}). Then, the complete action can be reconstructed for this illustrative case, where the evolution of the energy -density (\ref{2.12}) has been imposed in order to satisfy the usual continuity equation. An interesting remark comes from the fact that in the usual approach to modified gravity, as for instance $f(R)$ gravity,  pure dS solution (\ref{dSa}) does not admit the presence of a non-constant fluid, since the lhs of the field equations is constant, so the energy-matter sector can not depend on time. Nevertheless,  $f(R, T, P)$ gravity allows dS solutions in the presence of non-constant fluids due to the term $P=R_{\mu\nu}T^{\mu\nu}$ of the action. \\

Let us now consider the reconstruction of a theory capable of reproducing the Hubble parameter described by the $\Lambda$CDM model, but in the absence of a cosmological constant, a possibility already explored in $f(R)$ gravity (see \cite{gr-qc/0607118}). The following Hubble parameter is considered,
\be
H^2=\frac{\kappa^2}{3}\rho_0 a^{-3}+\frac{\Lambda}{3}\ .
\label{2.13}
\ee
Here, $\rho_0$ is the energy-density at the current time (with $a_0=1$), and $\Lambda$ is a constant. By expressing (\ref{2.13})
 in terms of the redshift $1+z=\frac{1}{a}$, and the cosmological parameters $\Omega_m^0=\frac{\rho_0}{3H_0^2/\kappa^2}$ and $\Omega_{\Lambda}^0=\frac{\Lambda}{3H_0^2}$, the Hubble parameter (\ref{2.13}) is rewritten as
 \be
H^2=H_0^2\Omega_m^0(1+z)^3+H_0^2\Omega_{\Lambda}^0\ .
\label{2.14}
\ee
Note that the model considered previously, $f(R,P)=R+f(P)$ would give rise to a completely different behavior of  the energy-density in comparison with (\ref{2.12}), except for the trivial solution $f(P)=-2\Lambda$, where GR is recovered. Nevertheless, by assuming the model $f(R, T, P)=R+f(P)+g(T)$, the evolution of the energy-density (\ref{2.12}) can be imposed  and the system results in a linear system of differential equations over $f$ and $g$, as was pointed above.  However, note that  the system (\ref{2.11}) can not be solved exactly in this case (and in general), but numerical calculations are required. By considering the energy-density evolution (\ref{2.12}) and the Hubble parameter (\ref{2.14}), the functions $P(z)$ and $T(z)$ take the form,
\be
P(z)=R_{\mu\nu}T^{\mu\nu}(z)=\frac{9H_0^4\Omega_m^0}{2\kappa^2}(\Omega_m^0(1+z)-2\Omega_{\Lambda})\ , \quad T(z)=-\frac{3H_0^2}{\kappa^2}\Omega_m^0 (1+z)^3\ ,
\label{2.15}
\ee
Then, the system of equations (\ref{2.11}) can be solved numerically for the Hubble parameter (\ref{2.14}), which leads to a set of functions for $F$ and $G$, defined in (\ref{2.16}). For an illustrative propose, some of the solutions are shown in fig.~1, which illustrates the kind of gravitational action compatible with the $\Lambda$CDM evolution  (\ref{2.14}). As a particular case, by setting all the derivatives of $F$ and $G$ to zero at $z=0$, but where $F(z=0)=G(z=0)=1$, the solution of the equations (\ref{2.11}) yields,
\be
f(R,T,R_{\mu\nu}T^{\mu\nu})= R+f(R_{\mu\nu}T^{\mu\nu})+g(T)=R-6H_0\Omega_{\Lambda}H_0^2=R-2\Lambda\ .
\label{2.18}
\ee
where we have used $\Omega_{\Lambda}^0=\frac{\Lambda}{3H_0^2}$. This is the trivial case that reduces the gravitational action to GR with a cosmological constant. However, there is an infinite number of solutions of the equations (\ref{2.11}) which keep the dependence both on $P$ as $T$.
\begin{figure*}[h!]
        \centering
\subfloat[]{\includegraphics[width=0.45\textwidth]{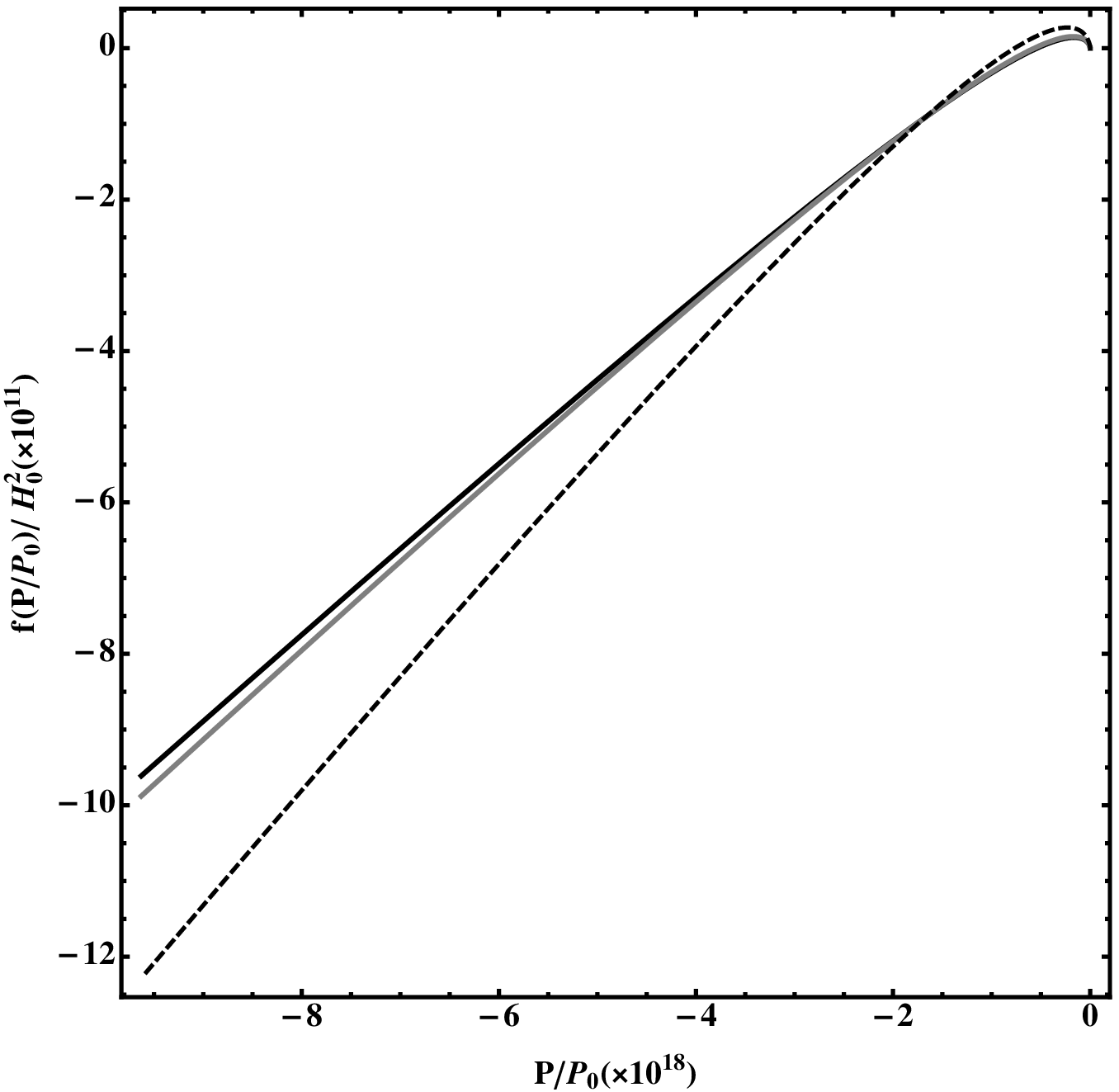}\label{fig_1}}\, \, \, \, \, \,
\subfloat[]{\includegraphics[width=0.45\textwidth]{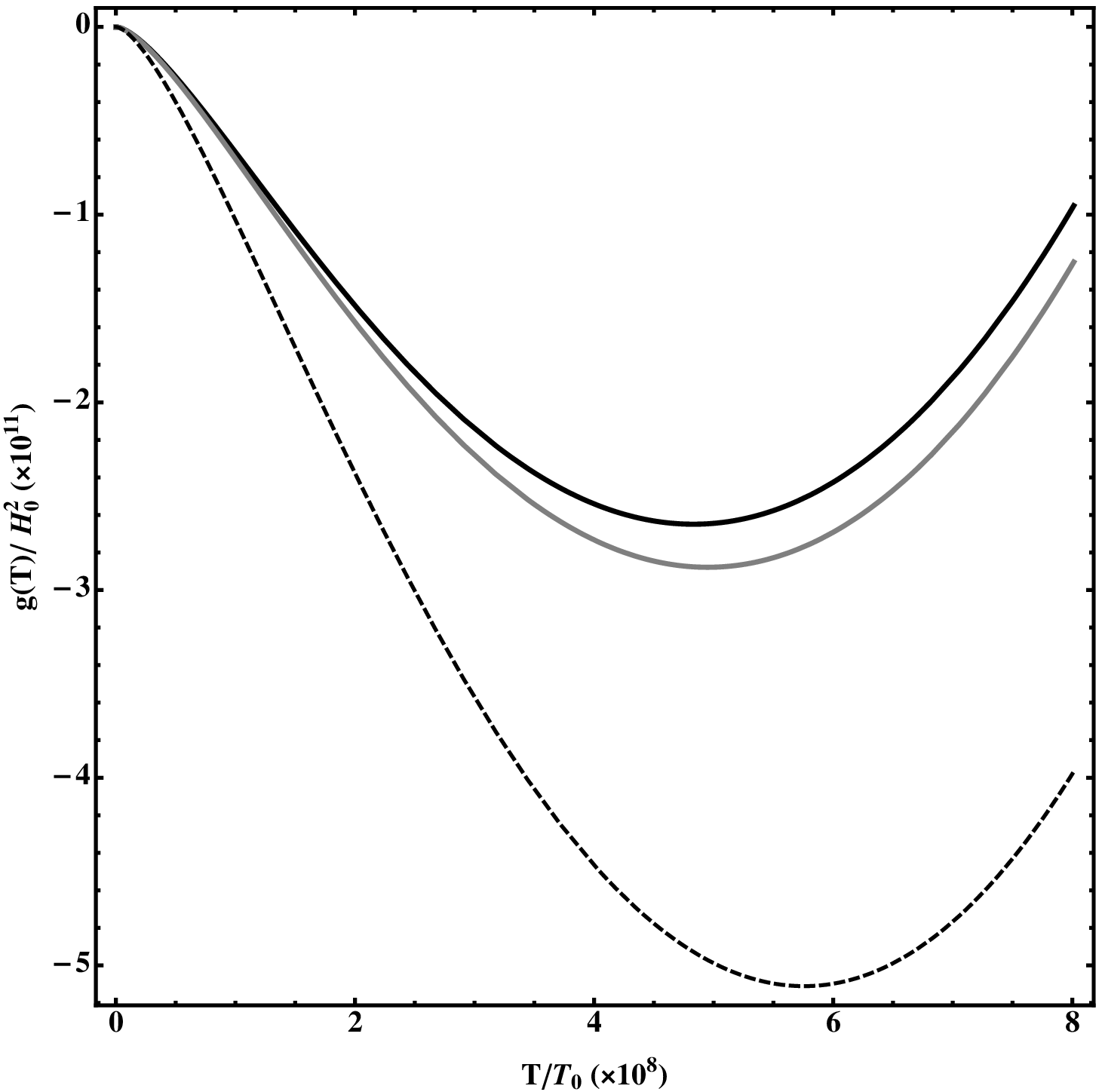}\label{fig_2}}
\caption{\footnotesize{The right panel corresponds to the function $F(P/P_0)$, whereas the left panel shows $G(T/T_0)$, both for the $\Lambda$CDM model. Each curve represents a set of initial conditions of the differential equations (\ref{2.11}), which have been set up at $z=0$, assuming values less than the unity for the derivatives $\{F^{(i)}(z), G^{(i)}(z)\}$ in order to be as close as possible to GR. }}
\end{figure*}

Using same technique one can reconstruct other types of evolution in the theory under discussion.

%%%%%%%%%%%%%%%%%%%%%%%%%%%%%%%%%%%%%
\section{Matter instability  of celestial body solutions}\label{instability}
%%%%%%%%%%%%%%%%%%%%%%%%%%%%%%%%%%%%%

This section is devoted to study  of so-called matter instability due to the modifications of the gravitational action. In the presence of gravitational objects, as the Sun or the Earth, tiny modifications of GR may make the system to become unstable in the interior solution,  a phenomena called the matter instability  (see Ref.~\cite{Dolgov:2003px}). This  problem has been already analyzed in $f(R)$ gravity, where it has been that matter instability can be avoided by an appropriate choice of the function that depends on the Ricci scalar (see Ref.~\cite{Nojiri:2003ft}).
Furthermore, it has been demonstrated that in viable $f(R)$ gravities such instability does not occur as well (for review, see \cite{0601213}).
 Nevertheless, the problem becomes much more complicated in the case of $f(R, T, R_{\mu\nu}T^{\mu\nu})$ gravity, since the field equations are very difficult to solve, even in the linear approximation, as shown below. \\

Let us consider for simplicity $f(R, T, R_{\mu\nu}T^{\mu\nu})=f_1(R)+f_2(T)+f_3(R_{\mu\nu}T^{\mu\nu})$. For a spherical body solution, the field equations have to be solved both in the exterior as well in the interior of the body. In the exterior, where $T=0$, the field equations (\ref{I7c}) reduce to the $f(R)$ gravity ones. The curvature $R_0$ becomes constant and is given by the solution of an algebraic equation, being $R_0=0$ in GR and in some $f(R)$ gravities. Nevertheless, the interior solution may differ substantially from GR, introducing corrections in the curvature that may lead to large instabilities. \\

For this analysis, the trace equation is considered, which can be easily obtained by multiplying the field equations (\ref{I7c}) by the metric $g^{\mu\nu}$
\be
Rf_{1_R}-2f+3\Box f_{1_R}+(T+\Theta)f_{2_T}+\frac{1}{2}\Box Tf_{3_T}+\nabla_{\alpha}\nabla_{\beta}T^{\alpha\beta}f_{3_P}+\Xi f_{3_P}=\kappa^2 T\ ,
\label{3.1}
\ee
where $P=R_{\mu\nu}T^{\mu\nu}$ as defined previously. Then, let us assume that $T=T_0$ describes correctly the interior of the celestial body, and according to GR the interior curvature gives
\be
R_0=-\kappa^2T_0\ ,
\label{3.2}
\ee
which is basically the trace equation of GR. In $f(R)$ gravity, the issue arises because the extra terms in the action may induce a perturbation on the curvature $\delta R$ that could grow very fast, giving rise to unstable spherical bodies. However, for some specific $f(R)$ functions, most of them called {\it viable} $f(R$ gravities, the problem can be circumvented, \cite{f(R)viable}. In $f(R, T, R_{\mu\nu}T^{\mu\nu})$ gravity, the problem becomes even worse, since the strong coupling among the curvature and the energy-momentum tensor may prevent $T_0$ being even a solution of the background equation. Nevertheless, by an appropriate $f_2(T)$ function, the energy-momentum tensor is able to behave as in GR, similarly as  shown in the previous section in the cosmological context, although the possible presence of a large perturbation $\delta R$, and additionally $\delta P$, or equivalently $T_0\delta R_{\mu\nu}$, may appear. Let us assume that $T=T_0=-\frac{R_0}{\kappa^2}$ is a solution of the background equation (\ref{3.1}), and a small perturbation in the curvature is considered
\be
R=R_0+\delta R\ , \quad P=P_0+\delta P\ .
\label{3.3}
\ee
Then, the equation (\ref{3.1}) at first linear order yields
\[
3f_1^{(2)}\Box \delta R-6\kappa^2 f_1^{(3)}g^{\alpha\beta}\nabla_{\alpha}T_0\nabla_{\beta} \delta R+\kappa^2\left[-T_0f_{1}^{(2)}-\frac{f_{1}^{(1)}}{\kappa^2}\right.
\]
\[
\left.-3f_{1}^{(3)}\Box T_0+3\kappa^2 f_{1}^{(4)}\nabla_{\alpha}T_0\nabla^{\alpha}T_0+P_0f_3^{(1)}\left(\mathcal{L}_m-\frac{1}{2}T_0\right)\right]\delta R+\frac{1}{2}T_0f_{3}^{(2)}\Box \delta P
\]
\[
+ \left[f_3^{(2)}\left(g^{\alpha\beta}\nabla_\alpha T_0+2\nabla_{\alpha}T^{\alpha\beta}_0+T^{\alpha\beta}_0\nabla_\alpha\right)+\left(T_0g^{\alpha\beta}+2T^{\alpha\beta}_0\right)f_3^{(3)}\nabla_\alpha P_0\right]\nabla_{\beta}\delta P
\]
\[
+\left[\Xi_0f_{3}^{(2)}+\frac{1}{2}\Box \left(f_{3}^{(2)}T_0\right)+\nabla_{\alpha}\nabla_{\beta}\left(f_{3}^{(2)}T^{\alpha\beta}_0\right)+2f_{3}^{(1)}P_0\right]\delta P=
\]
\be
=-3\tilde{\Box}f_{1}^{(1)}-\frac{1}{2}\left[\tilde{\Box}\left(T_0f^{(1)}_3\right)+2\tilde{\nabla}_{\alpha}\tilde{\nabla}_{\beta}\left(T^{\alpha\beta}f_{3}^{(1)}\right)\right]\ .
\label{3.4}
\ee
Here recall that $f_1=f_1(R)$ and $f_{3}=f_3(P)$, and the super-indexes denote derivatives with respect to $R$ and $P$, such as $f_{1}^{(i)}=d^if_1/dR^i$ and $f_{3}^{(i)}=d^if_3/dP^i$, whereas the tildes refer to covariant derivatives evaluated in the perturbed metric. Note that by setting $f_3=0$, the equation of the stability for $f(R)$ gravity is recovered Ref.~\cite{Nojiri:2003ft}. If the Hilbert-Einstein action is assumed, $f_1=R$, the solution of the equation (\ref{3.4}) gives $\delta R=0$, where no instabilities are produced, as expected in GR. Nevertheless, the equation (\ref{3.4}) is very complicated to be solved, even to get some qualitative information, since it involves not just perturbations of the Ricci scalar $\delta R$ but also of the Ricci tensor $\delta R_{\mu\nu}$ through $\delta P$. Since both the Ricci tensor and the Ricci scalar contain second derivatives of the metric, the equation (\ref{3.4}) should be rewritten in terms of the perturbed metric $\delta g_{\mu\nu}$, which gives rise to a more complicated equation. \\

Nevertheless, for an illustrative propose, let us consider the model of the previous section, where $f_1=R$, and the equation (\ref{3.4}) yields
\[
\frac{1}{2}T_0f_{3}^{(2)}\Box \delta P+ \left[f_3^{(2)}g^{\alpha\beta}\nabla_\alpha T_0+\left(T_0g^{\alpha\beta}+2T^{\alpha\beta}_0\right)f_3^{(3)}\nabla_\alpha P_0+2f_3^{(2)}\nabla_{\alpha}T^{\alpha\beta}_0\right.
\]
\[
\left.+f_3^{(2)}T^{\alpha\beta}_0\nabla_\alpha\right]\nabla_{\beta}\delta P+\left[\Xi_0f_{3}^{(2)}+\frac{1}{2}\Box \left(f_{3}^{(2)}T_0\right)+\nabla_{\alpha}\nabla_{\beta}\left(f_{3}^{(2)}T^{\alpha\beta}_0\right)+2f_{3}^{(1)}P_0\right]\delta P=
\]
\be
=-\frac{1}{2}\left[\tilde{\Box}\left(T_0f^{(1)}_3\right)+2\tilde{\nabla}_{\alpha}\tilde{\nabla}_{\beta}\left(T^{\alpha\beta}f_{3}^{(1)}\right)\right]+ \left[f_{1}^{(1)}-P_0f_3^{(1)}\left(\mathcal{L}_m-\frac{1}{2}T_0\right)\right]\delta R\ .
\label{3.5}
\ee
In addition, by considering a constant interior solution $T_0=$constant, and consequently $P_0=$constant, the equation (\ref{3.5}) is simplified,
\[
\Box \delta P+2\frac{T^{\alpha\beta}_0}{T_0}\nabla_\alpha\nabla_{\beta}\delta P+2\frac{\Xi_0f_{3}^{(2)}+2f_{3}^{(1)}P_0}{T_0f^{(2)}_3}\delta P=
\]
\be
= \frac{2}{T_0f^{(2)}_3}\left[f_{1}^{(1)}-P_0f_3^{(1)}\left(\mathcal{L}_m-\frac{1}{2}T_0\right)\right]\delta R\ .
\label{3.6}
\ee
A particular solution of (\ref{3.6}) is given by $\delta R=\delta P=0$, where no instability appears. Nevertheless, in general the behavior of the perturbation depends mainly on the term in front of $\delta P$ in (\ref{3.6}), where by imposing positivity on such term, $\frac{\Xi_0f_{3}^{(2)}+2f_{3}^{(1)}P_0}{T_0f^{(2)}_3}\geq0$, the general solution of the equation (\ref{3.6}) would approximately give rise to a damped oscillator, avoiding the appearance of a large instability in the interior of an spherical body. This example gives rise to a very restricted case, but  provides a qualitative description on the reconstruction of viable $f(R, T, R_{\mu\nu}T^{\mu\nu})$ gravities. \\

In addition, note that if $\delta P=T_0\delta R_{\mu\nu}$ is negligible in comparison with the first order term of the Ricci scalar $\delta R$, the dynamics of the perturbations would be ruled by the coefficient of the first term in (\ref{3.4}), so that if $3f_1^{(2)}\geq 0$, the DK instability may be avoided. Consequently, for the most general action $f(R, T, R_{\mu\nu}T^{\mu\nu})$, the DK instability may be avoided if $3f_{RR}-(T^{00}-\frac{1}{2}T)f_{PR}\geq 0$, as pointed out in Ref.~\cite{T.Harko}, provided that $\delta P=T_0\delta R_{\mu\nu}$ does not contribute significantly to the dynamics of the perturbations, a very strong approximation that is not satisfied in general. \\

Hence, a full analysis of the matter instability will probably reveal that a large instability is also common in this class of theories, which impose strong restrictions on the viability of the theory, but allows to reconstruct realistic $f(R, T, R_{\mu\nu}T^{\mu\nu})$ gravities.

%%%%%%%%%%%%%%%%%%%%%%%%
\section{Conclusions}
%%%%%%%%%%%%%%%%%%%%%%%%

In the present work, new version of modified gravity which includes   strong coupling of gravitational and matter fields, $R_{\mu\nu}T^{\mu\nu}$ has been studied.
The physical motivation for such theory comes from the covariant Ho\v{r}ava-like gravity with dynamical breaking of Lorentz invariance \cite{cov}. In fact, such a covariant power-counting renormalizable theory \cite{cov} represents simplest power-law $F(R,T,R_{\mu\nu}T^{\mu\nu})$ gravity.

Such modified gravity contains extra terms that allow to reconstruct viable cosmological evolution. In this sense, several cosmological solutions have been studied, and corresponding gravitational action is reconstructed. Nevertheless, the dynamics of the matter sector, a perfect fluid in the cases studied above, has to be fixed previously in order to guarantee the same evolution as in GR, or in other words to satisfy the continuity equation. Otherwise, a generic $f(R,T,R_{\mu\nu}T^{\mu\nu})$ may give a realistic Hubble parameter $H(z)$, but would give rise an anomalous behavior for baryonic, dark matter, and any other perfect fluid present in the equations, as shown in the first part of section \ref{Evolution}. In addition, this kind of theories allows to reproduce pure de Sitter universe in the presence of non-constant fluids what is usually a forbidden solution in GR, and  $f(R)$ or Gauss-Bonnet gravities.\\

Moreover, the $\Lambda$CDM universe can naturally occur in $f(R,T,R_{\mu\nu}T^{\mu\nu})$ gravity, where the usual evolution for dust matter, $\rho\propto (1+z)^3$, is set. This is shown in the second part of section \ref{Evolution} for the model described by the Hilbert-Einstein action plus corrections accounted by the functions $f(R_{\mu\nu}T^{\mu\nu})$ and $g(T)$ has been considered. The complexity of the equations does not allow to get an analytical and exact expression, but  using numerical results the behavior of $f(R_{\mu\nu}T^{\mu\nu})$ and $g(T)$ has been obtained, as shown in Fig.~1.\\

The last section has been devoted to the study of matter instability within these theories. The equation of the perturbations was obtained. It implies perturbations not just on the Ricci  scalar but also on the Ricci tensor. This fact makes complicated to get the analytical results since the equation should be reduced to a fourth order differential equation on the metric tensor by using the definitions of $R$ and $R_{\mu\nu}$. Nevertheless, by imposing
 some restrictions,  qualitative information has been obtained: it was shown that the reconstruction of $f_3(R_{\mu\nu}T^{\mu\nu})$ requires additional constraints in order to avoid matter instability. Moreover, assuming that $\delta P$ is negligible in comparison with $\delta R$, some constraints on the action are also obtained \cite{T.Harko}.  Then, some viable models, similarly as those in $f(R)$ gravity, capable of reproducing the realistic cosmological evolution could be reconstructed.\\

On the other hand, note that in general, the theory under consideration may contain ghosts due to higher-derivative terms in the action.
However, one can find some version of the theory which are free from ghosts. For instance, the theory with Lagrangian $L=F(R)+f_1(T)$  or some versions of the  theory with Lagrangian (3) (from Ref.{\cite{cov}) are ghost free.\\

Finally, It would be interesting to develop further the study of cosmological dynamics of the theory under discussion with attempts to simplify its Lagrangian formulation. This maybe probably achieved via the introduction of Lagrangian multipliers \cite{cov} and subsequent mapping of the theory to power-counting renormalizable covariant gravity with dynamical breaking of Lorentz invariance. In its turn this may indicate to nice ultraviolet properties of some $f(R,T,R_{\mu\nu}T^{\mu\nu})$ models and its relation with underlying quantum gravity.

\section*{Acknowledgments} 
The work by SDO is supported in part by MICINN(Spain), project FIS2010-15640, by AGAUR (Generalitat de Catalunya), contract 2009SGR-994 and by project 2.1839.2011 of MES (Russia).
DSG acknowledges support from the University of the Basque Country and the research project FIS2010-15640 (Spain), and also acknowledges the URC financial support from the University of Cape Town (South Africa). We are grateful to F.~Lobo and T.~Harko who expressed similar ideas, (see Ref.~\cite{T.Harko}), for useful discussions. We also thank A. de la Cruz-Dombriz for useful comments on the paper.

\end{document}